\documentclass[runningheads]{llncs}
\pdfoutput = 1

\usepackage{graphicx}
\usepackage{float}
\usepackage{array}
\usepackage{amsfonts}
\usepackage{amsmath}
\usepackage{amssymb}
\usepackage{setspace}
\usepackage{subfigure}
\usepackage{booktabs}
\usepackage{verbatim}
\usepackage{color}
\usepackage{bm}
\usepackage{balance}
\usepackage{multirow}
\usepackage{lipsum}
\usepackage[ruled,linesnumbered]{algorithm2e}

\usepackage{amsthm}
\usepackage{mathrsfs}

\newcommand{\hide}[1]{}
\newcommand{\atn}[1]{\textcolor{black}{#1}}
\newcommand{\zhangatn}[1]{\textcolor{black}{#1}}
\newcommand{\bingatn}[1]{\textcolor{black}{#1}}
\newcommand{\zry}[1]{\textcolor{black}{#1}}
\newcommand{\cjl}[1]{\textcolor{black}{#1}}

\newcommand{\reminder}[1]{{\textsf{\textcolor{blue}{[#1]}}}} 
\newcommand{\mkclean}{\renewcommand{\reminder}{\hide}}
\mkclean

\newcommand{\method}{\textsc{CubeFlow}\xspace}

\newcommand{\mb}{\mathbf}

\newtheorem{iproblem}{Informal Problem}

\newlength\myindent
\newcommand\blfootnote[1]{%
  \begingroup
  \renewcommand\thefootnote{}\footnote{#1}%
  \addtocounter{footnote}{-1}%
  \endgroup
}

\newcommand{\bit}{\begin{compactitem}}
\newcommand{\eit}{\end{compactitem}}
\newcommand{\ben}{\begin{compactenum}}
\newcommand{\een}{\end{compactenum}}

\usepackage[misc]{ifsym}
\usepackage{cite}



\graphicspath{{./figs/}}
\DeclareGraphicsExtensions{.pdf,.jpeg,.png}
\DeclareMathAlphabet\mathbfcal{OMS}{cmsy}{b}{n}
  
\begin{document}
\title{\method: Money Laundering Detection with Coupled Tensors}
\author{Xiaobing Sun$^{1,*}$,
Jiabao Zhang$^{1,2,*}$ 
Qiming Zhao$^{1,*}$,
Shenghua Liu$^{1,2,*}$\textsuperscript{(\Letter)},
Jinglei Chen$^3$,
Ruoyu Zhuang$^3$,
Huawei Shen$^{1,2}$,
Xueqi Cheng$^{1,2}$
}
\institute{
CAS Key Laboratory of Network Data Science and Technology, 
Insititute of Computing Technology, Chinese Academy of Sciences, China \\ 
\and University of Chinese Academy of Sciences, Beijing, China  \\
\and China Construction Bank Fintech, China \\
\email{xiaobingsun1999@gmail.com}, \email{zhangjiabao18@mails.ucas.edu.cn},
\email{qmzhao@cqu.edu.cn},
\email{liushenghua@ict.ac.cn},
\email{jl.chen.ray@gmail.com},
\email{zhuangruoyu@hotmail.com},
\email{\{shenhuawei,cxq\}@ict.ac.cn}
}
\authorrunning{X. Sun et al.}

 \maketitle      
\blfootnote{*Xiaobing Sun, Jiabao Zhang, and Qiming Zhao contribute equally as the first authors. Shenghua Liu is the corresponding author.The work was done when Xiaobing Sun and Qiming Zhao were visiting students at ICT CAS, who are separately from NanKai University and Chongqing University.}
\begin{abstract}
	\label{sec:abs}
	Money laundering (ML) is the behavior to conceal the source of money achieved by illegitimate activities, and always be a fast process involving frequent and chained transactions. 
How can we detect ML and fraudulent activity in large scale attributed transaction data (i.e.~tensors)?
Most existing methods detect dense blocks in a graph or a tensor, which do not consider the fact that money are frequently transferred through middle accounts. 
\method proposed in this paper is a scalable, flow-based approach to spot fraud from a mass of transactions by modeling them as two coupled tensors and applying a novel multi-attribute metric which can reveal the transfer chains accurately.
Extensive experiments show \method outperforms state-of-the-art baselines in ML behavior detection in both synthetic and real data.


\end{abstract}

\section{Introduction}
\label{sec:intro}
\begin{figure}[t]
\centering
\subfigure{
\includegraphics[height=1.3in]{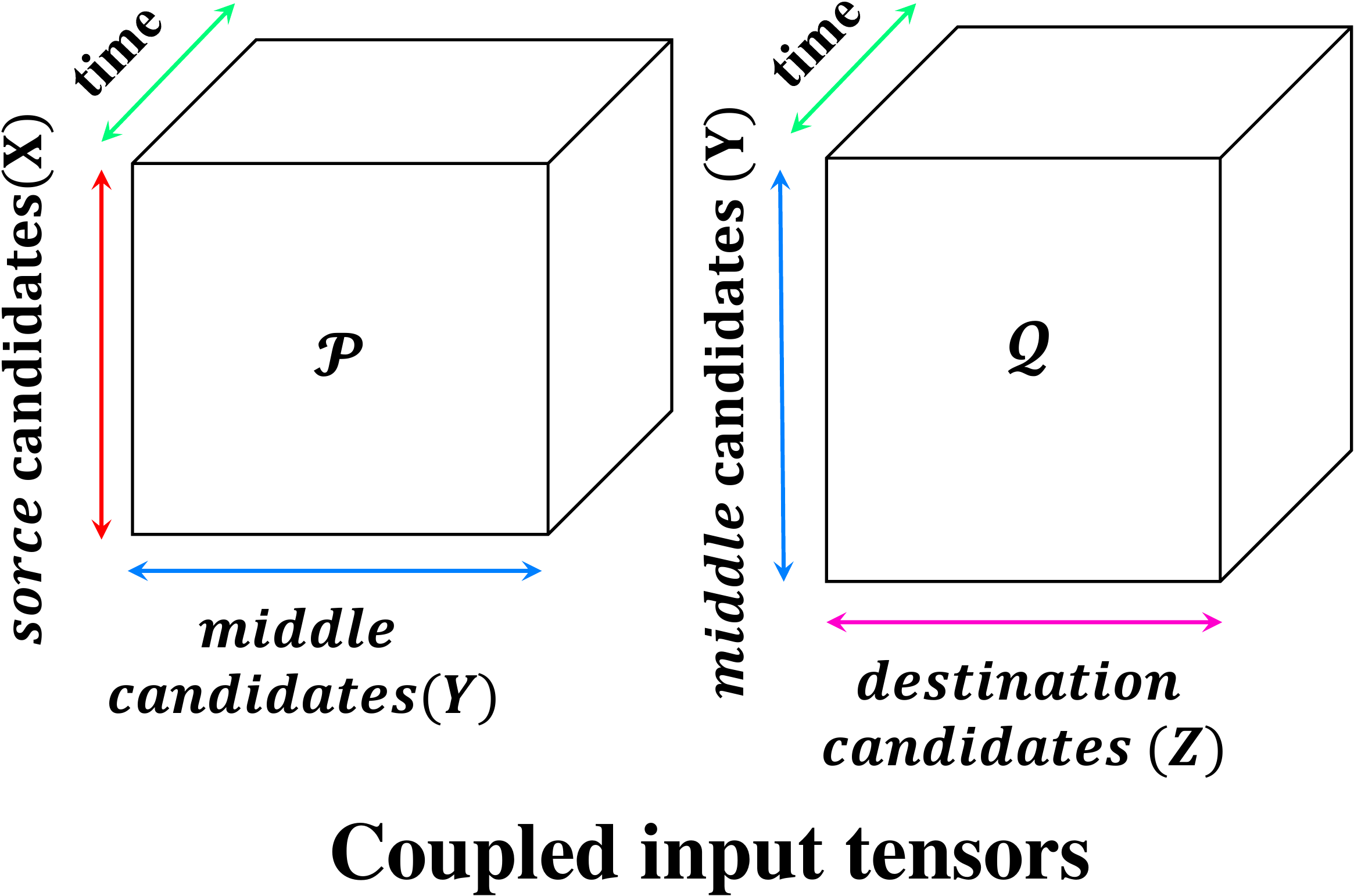}
}
\subfigure{
\includegraphics[height=1.3in]{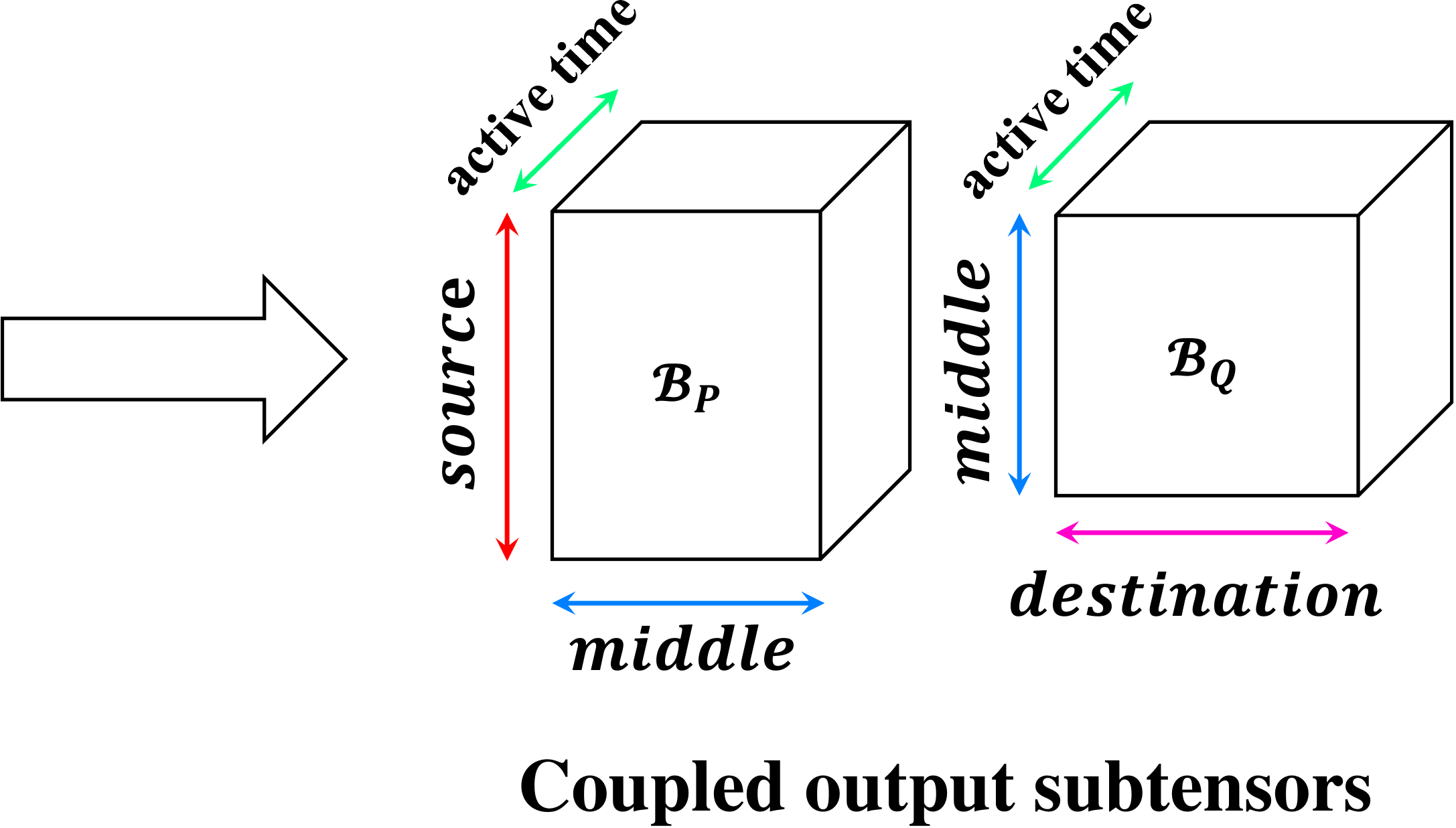}
}
\label{fig:basic_ML_behavior}
\caption{An example of ML detection. Two coupled input tensors indicate a money flow from $X$ to $Y$ to $Z$.  Modes $X, Y$ and $Z$ denote the candidates of source, middle and destination accounts. The purpose of detection is to find two dense coupled blocks in original coupled tensors, i.e., catching fraudsters involving in two-step ML activities.}
\end{figure}

Given a large amount of real-world transferring records, including a pair of accounts, some transaction attributes (e.g.~time, types), and volume of money, how can we 
detect money laundering (ML) accurately in a scalable way?
One of common ML processes disperses dirty money into 
different $source$ accounts, transfers them through many $middle$ accounts to $destination$ accounts for gathering in a fast way.
Thus the key problem for ML detection are:
\begin{iproblem}
\textbf{Given} a large amount of candidates of source, middle, and
destination accounts, and the transferring records, 
which can be formalized as two coupled tensors with 
entries of ($source~candidates$, $middle~candidates$, 
$time$, $\cdots$), 
and ($middle~candidates$, $destination~candidates$, $time$, $\cdots$),
how \textbf{to find} the accounts 
in such a ML process accurately and efficiently. 
\end{iproblem}
Fig.~\ref{fig:basic_ML_behavior} shows an example of ML detection with two coupled tensors indicating a flow from source to middle to destination accounts.
Those candidates can be pre-selected by existing feature-based models or in an empirical way. 
For example, in a bank, we can let source candidates simply be external accounts with more money transferring into the bank than out of the bank, destination candidates be the opposite ones, and middle candidates be the inner accounts.
\reminder{we need to bring the concept tensors, and say some comp with existing method.}
	
\reminder{Done: say some existing works. some words in sec 3 should move to intro}
Most existing dense subtensor detection methods~\cite{shin2017d,Jiang2016c,shin2016m} have been used for tensor fraud detection, but only can deal with one independent tensor. 
Therefore, they exploit exactly single-step transfers, but do not account for ``transfer chains". 
Such methods based on graph's density~\cite{hooi2016fraudar,liu2017holoscope,prakash2010eigenspokes}, have the same problem and even not be able to leverage multi-attributes. 
Although, FlowScope \cite{Li2020fs} designed for dense and multi-step flow, it fails to take into account some important properties (e.g.~time) because of the limits by the graph.

Therefore, we propose \method, a ML detection method with coupled tensors. \method not only considers the flow of funds (from sources, through middle accounts, to destinations), but also can combine some attributes, 
such as transferring time to model
fraudsters' highly frequent transfers. 
We define a novel multi-attribute metric for fraudulent
transferring flows of ML. 
\method considers the \atn{suspicious in-and-out balance} for middle accounts within short time intervals and detects the chain of fraudulent transfers accurately. The experiments on real-world datasets show that \method detects various adversarial injections and real ML fraudsters both with high accuracy and robustness.

In summary, the main advantages of our work are:

$\bullet$ \textbf{Multi-attribute metric for money-laundering flow:} 
We propose a novel multi-attribute metric for detecting dense transferring flows in coupled
tensors, which measures the anomalousness of
typical two-step laundering with suspicious in-and-out balance for middle accounts within 
many short time intervals.
\reminder{Roughly Done: suppose the reader did not know FlowScope. we need to say both: money are frequently transferred, and little money left in middle accounts. Note no explain for fast in and fast out yet.}



$\bullet$ \textbf{Effectiveness and robustness:} 
\reminder{Done: summarize the experiments. reduce money, and increase accounts}
\method outperforms baselines under various injection densities by descending the amount of money or ascending \# of accounts on real-world datasets. And \method shows its robustness in
different proportions of accounts with better accuracy. \reminder{As for real ML activities in a bank, our \method can detect them in a 
reasonable accuracy with about 30\% money are transferred.
}


$\bullet$ \textbf{Scalability:} \method is scalable, with near-linear time complexity in the number of transferring records.

Our code and processed data are publicly available for reproducibility~\footnote{\url{https://github.com/BGT-M/spartan2-tutorials/blob/master/CubeFlow.ipynb}}.

\section{Related Work}
\label{sec:relate}
Our work pulls from two main different fields of research: \zhangatn{(i) domain-specific ML detection methods; and (ii) general anomaly detection methods in graphs and tensors. }

\textbf{Money laundering detection.}
\zry{The most classical approaches for Anti-ML are those of rule based classification relying heavily on expertise.}
Khan et al. \cite{khan2013bayesian} used Bayesian network designed \zry{with  guidance of the rules to assign risk scores to transactions.}
\zry{The system proposed by \cite{khanuja2014forensic} monitored ongoing transactions and assessed their degree of anomaly.} 
However, rule based algorithms are easy to be evaded by fraudsters. 
To involve more attributes and handle the high-dimensional data, machine learning models such as SVM \cite{tang2005developing}, decision trees \cite{wang2007money} and neural networks \cite{lv2008rbf} are applied, while these methods are focused on
isolated transaction level. 
Stavarache et al. \cite{2019Exploring} proposed a deep learning based method trained for Anti-ML tasks using customer-to-customer
relations. 
However, these algorithms detect the ML activities in supervised or semi-supervised manners, suffering from imbalanced class and lack of adaptability and interpretability. 


\textbf{General-purpose anomaly detection in graphs and tensors.}
\zhangatn{Graphs (i.e.~tensors) provide a powerful mechanism to capture interrelated associations between data objects~\cite{akoglu2015graph}, and there have been many
graph-based techniques developed for discovering
structural anomalies. }
SpokEn~\cite{prakash2010eigenspokes} studied patterns in eigenvectors, and was  applied for anomaly detection in ~\cite{jiang2014inferring} later.
CatchSync~\cite{jiang2014catchsync} exploited two of the tell-tale signs created by fraudsters.
And many existing methods rely on graph (i.e.~tensor)'s density, e.g., Fraudar~\cite{hooi2016fraudar} proposed a suspiciousness measure on the density,  HoloScope~\cite{liu2017holoscope, 8494803} considered temporal spikes and hyperbolic topology and SpecGreedy~\cite{feng2020specgreedy} proposed a unified framework based on the graph spectral properties.
D-Cube~\cite{shin2017d}, M-Zoom~\cite{shin2016m} and CrossSpot~\cite{Jiang2016c} adopted greedy approximation algorithms to detect dense subtensors, while CP Decomposition (CPD)~\cite{kolda2009t}
focused on tensor decomposition methods.
However, \zhangatn{these methods are designed for general-purpose anomaly detection tasks, which not take the flow across multiple nodes into account.}

\section{PROBLEM FORMULATION}
\label{sec:problem}
\begin{table}[t]
    \centering
    \footnotesize
    \caption{Notations and symbols}
    \begin{tabular}{>{\centering\arraybackslash}p{0.27\linewidth}|p{0.72\linewidth}}
    \hline
    \textbf{Symbol} & \textbf{Definition} \\
    \hline
    ${\mathbfcal{P}}(X, Y, A_3, \ldots, A_N, V)$ & Relation representing the tensor which stands money trans from $X$ to $Y$ \\
    ${\mathbfcal{Q}}(Y, Z, A_3, \ldots, A_N, V)$ & Relation representing the tensor which stands money trans from $Y$ to $Z$ \\
    $N$ & Number of mode attributes in $\mathbfcal{P}$ (or $\mathbfcal{Q}$) \\
    $A_n$ & $n$-th mode attribute name in $\mathbfcal{P}$ (or $\mathbfcal{Q}$) \\
    $V$ & measure attribute (e.g.~money) in ${\mathbfcal{P}}$ (or $\mathbfcal{Q}$)\\
    ${\mathbfcal{B}}_{\bf{P}}$ (or ${\mathbfcal{B}}_{\bf{Q}}$) & a block(i.e.~ subtensor) in $\mathbfcal{P}$ (or $\mathbfcal{Q}$) \\
    ${\bf{P}}_{x}$, ${\bf{P}}_{y}$, ${\bf{Q}}_{z}$,${\bf{P}}_{a_n}$  & set of distinct values of $X,Y,Z,A_n$ in $\mathbfcal{P}$ (or $\mathbfcal{Q}$) \\
    ${\bf{B}}_x$,${\bf{B}}_y$, ${\bf{B}}_z$, ${\bf{B}}_{a_n}$  & set of distinct values of $X,Y,Z,A_n$ in ${\mathbfcal{B}}_{\bf{P}}$ (or ${\mathbfcal{B}}_{\bf{Q}}$) \\
   
    $M_{x,y,a_3,\ldots,a_N}({\mathbfcal{B}}_{\bf{P}})$ & attribute-value mass of $(x,y,a_3,\ldots,a_N)$ in $(X,Y,A_3,\ldots,A_N)$ \\
    $M_{y,z,a_3,\ldots,a_N}({\mathbfcal{B}}_{\bf{Q}})$ & attribute-value mass of $(y,z,a_3,\ldots,a_N)$ in $(Y,Z,A_3,\ldots,A_N)$ \\
    $\omega_{i}$ & Weighted assined to a node in priority tree \\
    $g({\mathbfcal{B}}_{\bf{P}},{\mathbfcal{B}}_{\bf{Q}})$ & Metric of ML anomalousness\\
    $[3,N]$ & $\{3,4,\ldots,N\}$\\
    \hline
    \end{tabular}
    \label{tab:symboldef}
\end{table}


\cjl{In general money laundering (ML) scenario, fraudsters transfer money from source accounts to destination accounts through several middle accounts in order to cover up the true source of funds. Here, we summarize three typical characteristics of money laundering:}

\textbf{Density:} \cjl{In ML activities, a high volume of funds needs to be transferred from source to destination accounts with limited number of middle accounts. Due to the risk of detection, fraudsters tend to use shorter time and fewer trading channels in the process of ML which will create a high-volume and dense subtensor of transfers.}

\textbf{Zero out middle accounts:} 
\cjl{The role of middle accounts can be regarded as a bridge in ML: only a small amount of balance will be kept in these accounts for the sake of camouflage, most of the received money will be transferred out. This is because the less balances retained, the less losses will be incurred if these accounts are detected or frozen.}

\textbf{Fast in and Fast out:}
\cjl{To reduce banks' attention, ``dirty money" is always divided into multiple parts and transferred through the middle accounts one by one. The ``transfer", which means a part of fund is transferred in and out of a middle account, is usually done within a very short time interval. This is because the sooner the transfer is done, the more benefits fraudsters will get.}


    Algorithms which focus on individual transfers, e.g. feature-based	approaches, can be easily evaded by adversaries by keeping each individual transfer looks normal. 
	Instead, our goal is to detect dense blocks in tensors composed of source, middle, destination accounts and other multi-attributes, as follows:

\begin{iproblem}[ML detection with coupling tensor]
    \textbf{Given} two money transfer tensors ${\mathbfcal{P}}(X, Y, A_3, \ldots, A_N, V)$ and ${\mathbfcal{Q}}(Y, Z, A_3, \ldots, A_N, V)$, \bingatn{with attributes of source, middle and destination candidates as $X, Y, Z$}, other coupling attributes (e.g.~time) as $A_n$, and a nonnegative \bingatn{measure} attribute (e.g.~\atn{volume of money}) as $V$.\\
    \indent \textbf{Find:} two dense blocks (i.e.~subtensor) of $\mathbfcal{P}$ and $\mathbfcal{Q}$.\\
    \indent \textbf{Such that:} \\
    \indent \textbf{-} it maximizes density.\\
    \indent \textbf{-} for each middle account, the money transfers satisfy zero-out and fast-in-and-fast-out characteristics. \\
    
\end{iproblem}

\cjl{Symbols used in the paper are listed in Table \ref{tab:symboldef}. As common in other literature, we denote tensors and modes of tensors by boldface calligraphic letters(e.g.~${\mathbfcal{P}}$) and capital letters(e.g.~$X,Y,Z,{A_n}$) individually. For the possible values of different modes, boldface uppercase letters (e.g.~${{\bf{P}}_{x}}, {{\bf{P}}_{y}}, {{\bf{Q}}_{z}}, {{\bf{P}}_{a_n}}$) are used  in this paper. Since ${\mathbfcal{P}}(X, Y, A_3, \ldots, A_N, V)$ and ${\mathbfcal{Q}}(Y, Z, A_3, \ldots, A_N, V)$   are coupled
tensors sharing the same sets of modes ${Y,A_n}$, we have ${\bf{P}}_{y} =  {\bf{Q}}_{y}$ and ${\bf{P}}_{a_n}$ = ${\bf{Q}}_{a_n}$.
Our targets, the dense blocks (i.e.~subtensor) of $\mathbfcal{P}$ and $\mathbfcal{Q}$, are represented by ${\mathbfcal{B}}_{\bf{P}}$ and ${\mathbfcal{B}}_{\bf{Q}}$. Similarly, the mode's possible values in these blocks are written as ${{\bf{B}}_{x}}, {{\bf{B}}_{y}}, {{\bf{B}}_{z}}, {{\bf{B}}_{a_n}}$. An entry $(x,y,a_3,\ldots,a_N)$  indicates that account ${x}$ transfers money to account ${y}$ when other modes are equal to $a_3,\ldots,a_N$ (e.g. during $a_3$ time-bin), and $M_{x,y,a_3,\ldots,a_N}({\mathbfcal{B}}_{\bf{P}}) $ is the total amount of money on the subtensor.}

\section{Proposed Method}
\label{sec:meth}
\subsection{Proposed Metric}

\cjl{First,  we give the concept of fiber: \emph{A fiber of a tensor ${\mathbfcal{P}}$ is a vector obtained by fixing all but one ${\mathbfcal{P}}$'s indices}. For example, in ML process with $A_3$ representing transaction timestamp, total money transferred from source accounts ${X}$ into a middle account ${y}$ at time-bin ${a_3}$ is the mass of fiber ${\mathbfcal{P}}(X, y, a_3, V)$ which can be denoted by ${M_{:,y,a_3}({\mathbfcal{B}}_{\bf{P}})}$, while total money out of the middle account can be denoted by ${M_{y,:,a_3}({\mathbfcal{B}}_{\bf{Q}})}$.}

In general form, we can define the minimum and maximum value between total amount of money transferred into and out of a middle account ${y\in{{\bf{B}}_y}}$ with other attributes equal to $a_3\in{{\bf{B}}_{a_3}}, \ldots, a_N\in{{\bf{B}}_{a_N}}$:
\begin{equation}
    {f_{y,a_3,\ldots,a_N}}({\mathbfcal{B}}_{\bf{P}},{\mathbfcal{B}}_{\bf{Q}}) = min\{M_{:,y,a_3,\ldots,a_N}({\mathbfcal{B}}_{\bf{P}}),  M_{y,:,a_3,\ldots,a_N}({\mathbfcal{B}}_{\bf{Q}})\}
\end{equation}

\begin{equation}
    {q_{y,a_3,\ldots,a_N}}({\mathbfcal{B}}_{\bf{P}},{\mathbfcal{B}}_{\bf{Q}}) = max\{M_{:,y,a_3,\ldots,a_N}({\mathbfcal{B}}_{\bf{P}}),  M_{y,:,a_3,\ldots,a_N}({\mathbfcal{B}}_{\bf{Q}})\}
\end{equation}

Then we can define the difference between the maximum and minimum value:
\begin{equation}
    {r_{y,a_3,\ldots,a_N}}({\mathbfcal{B}}_{\bf{P}},{\mathbfcal{B}}_{\bf{Q}}) = {q_{y,a_3,\ldots,a_N}}({\mathbfcal{B}}_{\bf{P}},{\mathbfcal{B}}_{\bf{Q}}) - {f_{y,a_3,\ldots,a_N}}({\mathbfcal{B}}_{\bf{P}},{\mathbfcal{B}}_{\bf{Q}})
\end{equation}

Next, our ML metric is defined as follows for spotting multi-attribute money-laundering flow:
\begin{definition}
(Anomalousness of coupled blocks of ML)
The anomalousness of a flow from a set of nodes ${\bf{B}}_x$,
through the inner accounts ${\bf{B}}_y$, to another subset ${\bf{B}}_z$, where other attribute values are ${a_n} \in {\bf{B}}_{a_n} $:
\begin{equation}
    \label{metricdef}
    \begin{aligned}
        g({\mathbfcal{B}}_{\bf{P}},{\mathbfcal{B}}_{\bf{Q}})
        & = \frac{{\sum_{{y\in{{\bf{B}}_y},{{a_i}\in{{\bf{B}}_{a_i}}}}}} \big({(1-\alpha){f_{y,a_3,\ldots,a_N}}({\mathbfcal{B}}_{\bf{P}},{\mathbfcal{B}}_{\bf{Q}}) - \alpha\cdot{r_{y,a_3,\ldots,a_N}}({\mathbfcal{B}}_{\bf{P}},{\mathbfcal{B}}_{\bf{Q}})}\big)}{{\sum_{i=3}^{N}(|{\bf{B}}_{a_i}|)} + {|{\bf{B}}_x|} + {|{\bf{B}}_y|} + {|{\bf{B}}_z|}} \\
        & = \frac{{\sum_{{y\in{{\bf{B}}_y},{{a_i}\in{{\bf{B}}_{a_i}}}}}} \big({f_{y,a_3,\ldots,a_N}}({\mathbfcal{B}}_{\bf{P}},{\mathbfcal{B}}_{\bf{Q}}) - \alpha{q_{y,a_3,\ldots,a_N}}({\mathbfcal{B}}_{\bf{P}},{\mathbfcal{B}}_{\bf{Q}})\big)}{{\sum_{i=3}^{N}(|{\bf{B}}_{a_i}|)} + {|{\bf{B}}_x|} + {|{\bf{B}}_y|} + {|{\bf{B}}_z|}} 
    \end{aligned}
\end{equation}
\end{definition}
\reminder{Done: How about keeping the sum of $B_{a_i}$, and in algorithm we did a approximation of $\mb I$?}

Intuitively, ${{f_{y,a_3,\ldots,a_N}}({\mathbfcal{B}}_{\bf{P}})}$ is the maximum possible flow that could go through middle account $y\in{{\bf{B}}_y}$ when other attributes are ${a_n} \in {\bf{B}}_{a_n} $. ${r_{y,a_3,\ldots,a_N}}({\mathbfcal{B}}_{\bf{P}},{\mathbfcal{B}}_{\bf{Q}})$ is the absolute value of ``remaining money" in account $y$ after transfer, i.e., retention or deficit, which can be regarded as a penalty for ML, \cjl{ since fraudsters prefer to keep small account balance at any situations.} 
When we set $A_3$ as time dimension, we consider the ``remaining money" in each time bin which will catch the trait of  fast in and fast out during ML.
We define $\alpha$ as the coefficient of imbalance cost rate in the range of 0 to 1.
\subsection{Proposed Algorithm: \method}
We use a near-greedy algorithm \method, to find two dense blocks ${\mathbfcal{B}}_{\bf{P}}$ and ${\mathbfcal{B}}_{\bf{Q}}$ maximizing the objective $g({\mathbfcal{B}}_{\bf{P}},{\mathbfcal{B}}_{\bf{Q}})$ in (\ref{metricdef}).

To develop an efficient algorithm for our metric, we unfold the tensor ${\mathbfcal{P}}$ on mode-$X$ and ${\mathbfcal{Q}}$ on mode-$Z$. For example, a tensor unfolding of ${{\mathbfcal{P}}\in{{\mathbb{R}}^{|{\bf{B}}_x|\times|{\bf{B}}_y|\times|{\bf{B}}_{a_3}|}}}$  on mode-$X$ will produce a $|{\bf{B}}_x|\times(|{\bf{B}}_y|\times{|\bf{B}}_{a_3}|)$ matrix.

For clarity, we define the index set $\bf{I}$, whose size equals to the number of columns of matrix:  
\begin{equation}
    {\bf{I}} = {{\bf{B}}_y}\Join{{\bf{B}}_{a_{3}}}\Join\ldots\Join{{\bf{B}}_{a_{N}}}
\end{equation} 
where $\Join$ denotes Cartesian product. Therefore, the denominator of \eqref{metricdef} can be approximated by ${{|{\bf{B}}_x|} + {|{\bf{I}}|} + {|{\bf{B}}_z|}}$.

First, we build a priority tree for \atn{entries} in ${\mathbfcal{B}}_{\bf{P}}$ and ${\mathbfcal{B}}_{\bf{Q}}$. The weight (ie. priority) assigned to \bingatn{index} $i$ is defined as:
\begin{equation}
    \label{weightdef}
    {\omega_i}({\mathbfcal{B}}_{\bf{P}},{\mathbfcal{B}}_{\bf{Q}})=\left\{
        \begin{array}{lcl}
        {f_{i}}({\mathbfcal{B}}_{\bf{P}},{\mathbfcal{B}}_{\bf{Q}}) - \alpha{q_{i}}({\mathbfcal{B}}_{\bf{P}},{\mathbfcal{B}}_{\bf{Q}}), & & {\text{if } i\in {\bf{I}}}\\
        M_{i,:,:,\ldots,:}({\mathbfcal{B}}_{\bf{P}}), & & {\text{if } i\in{{\bf{B}}_x}}\\
       M_{:,i,:,\ldots,:}({\mathbfcal{B}}_{\bf{Q}}), & & {\text{if } i\in{{\bf{B}}_z}}\\
        \end{array} \right.
\end{equation}

The algorithm is described in Alg~\ref{algorithm_code}. 
After building the priority tree, we perform the near greedy optimization: block ${\mathbfcal{B}}_{\bf{P}}$ and ${\mathbfcal{B}}_{\bf{Q}}$ start with whole tensor ${\mathbfcal{P}}$ and ${\mathbfcal{Q}}$. 
Let we denote ${{\bf{I}}\cup{\bf{B}}_x\cup{\bf{B}}_z}$ as ${\bf{S}}$. In every iteration, we remove the node $v$ in ${\bf{S}}$ with minimum weight in the tree, approximately maximizing objective \eqref{metricdef}; 
and then we update the weight of all its neighbors. 
The iteration is repeated until one of node sets ${\bf{B}}_x, {\bf{B}}_z, {\bf{I}}$ is empty.
Finally, two dense blocks ${\hat{\mathbfcal{B}}_{\bf{P}}},{\hat{\mathbfcal{B}}_{\bf{Q}}}$ that we have seen with the largest value $g({\hat{\mathbfcal{B}}_{\bf{P}}},{\hat{\mathbfcal{B}}_{\bf{Q}}})$ are returned.

\begin{algorithm}[H]
    \caption{\method}
    \small
    \label{algorithm_code}
    \LinesNumbered 
    \KwIn{relation ${\mathbfcal{P}}$, relation ${\mathbfcal{Q}}$}
    \KwOut{dense block ${\mathbfcal{B}}_{\bf{P}}$, dense block ${\mathbfcal{B}}_{\bf{Q}}$}
    ${\mathbfcal{B}}_{\bf{P}}\leftarrow{\mathbfcal{P}}$, ${\mathbfcal{B}}_{\bf{Q}}\leftarrow{\mathbfcal{Q}}$\;
    ${\bf{S}} \leftarrow {{\bf{I}}\cup{\bf{B}}_x\cup{\bf{B}}_z}$\;
    $\omega_i \leftarrow$ calculate node weight as Eq. (\ref{weightdef}) \;
    $T \leftarrow$ build priority tree for ${\mathbfcal{B}}_{\bf{P}}$ and ${\mathbfcal{B}}_{\bf{Q}}$ with ${\omega_i}({\mathbfcal{B}}_{\bf{P}},{\mathbfcal{B}}_{\bf{Q}})$ \;
     \While{${\bf{B}}_x, {\bf{B}}_z$ and ${\bf{I}}$ is not empty}{
        $v \leftarrow$ find the minimum weighted node in T\;
        ${\bf{S}} \leftarrow {\bf{S}} \backslash \{v\}$\;
        update priorities in $T$ for all neighbors of $v$\;
        $ g({\mathbfcal{B}}_{\bf{P}},{\mathbfcal{B}}_{\bf{Q}}) \leftarrow$ calculate as Eq. (\ref{metricdef})\;
    }
    return ${\hat {\mathbfcal{B}}_{\bf{P}}},{\hat{\mathbfcal{B}}_{\bf{Q}}}$ that maximizes $g({\mathbfcal{B}}_{\bf{P}},{\mathbfcal{B}}_{\bf{Q}})$ seen during the loop.
\end{algorithm}

\section{Experiments}
\label{sec:exp}

We design experiments to answer the following questions:

\begin{itemize}
	\item \textbf{Q1. Effectiveness:} How early and accurate does our method detect synthetic ML behavior comparing to the baselines?
	\item \textbf{Q2. Performance on real-world data:} How early and accurate does our \method detect real-world ML activity comparing to the baselines?
	\item \textbf{Q3. Performance on $4$-mode tensor:} How accurate does \method compare to the baselines dealing with multi-mode data?
	\item \textbf{Q4. Scalability:}  Does our method scale linearly with the number of edges?
\end{itemize}

\begin{table}[thp]
    \centering
    \small
    \caption{Dataset Description}
    \begin{tabular}{lll}
        \toprule
        \multicolumn{1}{c|}{Name} & \multicolumn{1}{c|}{Volume} & 
        \multicolumn{1}{c}{\# Tuples} \\
        \midrule
        \multicolumn{3}{l}{3-mode bank transfer record (\underline{from\_acct}, \underline{to\_acct}, \underline{time}, money)}\\
        \midrule
        \multicolumn{1}{c|}{\multirow{2}{*}{CBank~}} & 
        \multicolumn{1}{c|}{~$491295\times561699\times576$~} & \multicolumn{1}{c}{$2.94M$} \\
        \multicolumn{1}{c|}{} &
        \multicolumn{1}{c|}{~~$561699\times1370249\times576$~~} &
        \multicolumn{1}{c}{$2.60M$} \\
        \midrule
        \multicolumn{1}{c|}{\multirow{2}{*}{CFD-3~}} & 
        \multicolumn{1}{c|}{$2030\times2351\times728$} & \multicolumn{1}{c}{$0.12M$} \\
        \multicolumn{1}{c|}{} &
        \multicolumn{1}{c|}{$2351\times7001\times730$} &
        \multicolumn{1}{c}{$0.27M$} \\
        \midrule
        \multicolumn{3}{l}{4-mode bank transfer record (\underline{from\_acct}, \underline{to\_acct}, \underline{time}, \underline{k\_symbol}, money)}\\
        \midrule
        \multicolumn{1}{c|}{\multirow{2}{*}{CFD-4~}} & 
        \multicolumn{1}{c|}{$2030\times2351\times728\times7$} & \multicolumn{1}{c}{$0.12M$} \\
        \multicolumn{1}{c|}{} &
        \multicolumn{1}{c|}{$2351\times7001\times730\times8$} &
        \multicolumn{1}{c}{$0.27M$} \\
    \bottomrule
    \end{tabular}
    \label{tab:data}
\end{table}

\subsection{Experimental Setting}
\textbf{Machine:} We ran all experiments on a machine with 2.7GHZ Intel Xeon E7-8837 CPUs and 512GB memory.

\textbf{Data:}
Table~\ref{tab:data} lists data used in our paper. CBank data is a real-world transferring data from an anonymous bank under an NDA agreement. Czech Financial Data (CFD) is an anonymous transferring data of Czech bank released for Discovery Challenge in \cite{senond_dataset}.
We model CFD data as two $3$-mode tensors consisting of entries ($a_1$, $a_2$, $t$, $m$), which means that account $a_1$ transfers the amount of money, $m$,  to account $a_2$ at time $t$. Specifically, we divide the account whose money transferring into it is much larger than out of the account into $X$, on the contrary, into $Z$, and the rest into $Y$.
Note that it does not mean that $X,Y,Z$ have to be disjoint, while this preprocessing helps speed up our algorithm.
We also model CFD data as two $4$-mode tensors having an additional dimension \emph{k\_Symbol}~(characterization of transaction, e.g., insurance payment and payment of statement). And we call two CFD data as CFD-$3$ and CFD-$4$ resp.

\textbf{Implementations:}
We implement \method in Python, CP Decomposition(CPD)\cite{kolda2009t} in Matlab and run the open source code of D-Cube~\cite{shin2017d}, M-Zoom~\cite{shin2016m}and CrossSpot~\cite{Jiang2016c}. We use the sparse tensor format for efficient space utility. Besides, the length of time bins of CBank and CFD are $20$ minutes and $3$ days respectively, and the value of $\alpha$ is $0.8$ as default.

\subsection{Q1.Effectiveness}
To verify the effectiveness of \method, we inject ML activities as follows: fraudulent accounts are randomly chosen as the tripartite groups, denoted by $B_x$, $B_y$ and $B_z$.
The fraudulent edges between each group are randomly generated with probability $p$. We use Dirichlet distribution (the value of scaling parameter is $100$) to generate the amount of money for each edge. And for each account in $B_y$, the amount of money received from $B_x$ and that of transferred to $B_z$ are almost the same. Actually, we can regard the remaining money of accounts in $B_y$ as camouflage, with amount of money conforms to a random distribution ranging from $0$ to $100000$ (less than $1\%$ of injected amount of money). 
To satisfy the trait ``Fast in and Fast out'', we randomly choose the time from one time bin for all edges connected with the same middle account.

\begin{figure}[t]
    \centering
    \begin{minipage}{0.8\linewidth}
        \includegraphics[width=\textwidth]{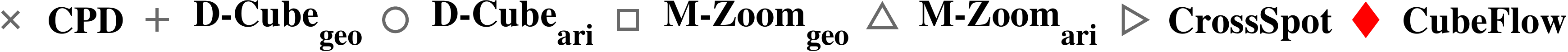}
    \end{minipage}
    \vfill 
    \subfigure[descending amount of money]{
        \begin{minipage}{0.36\linewidth}
            \includegraphics[width=\textwidth]{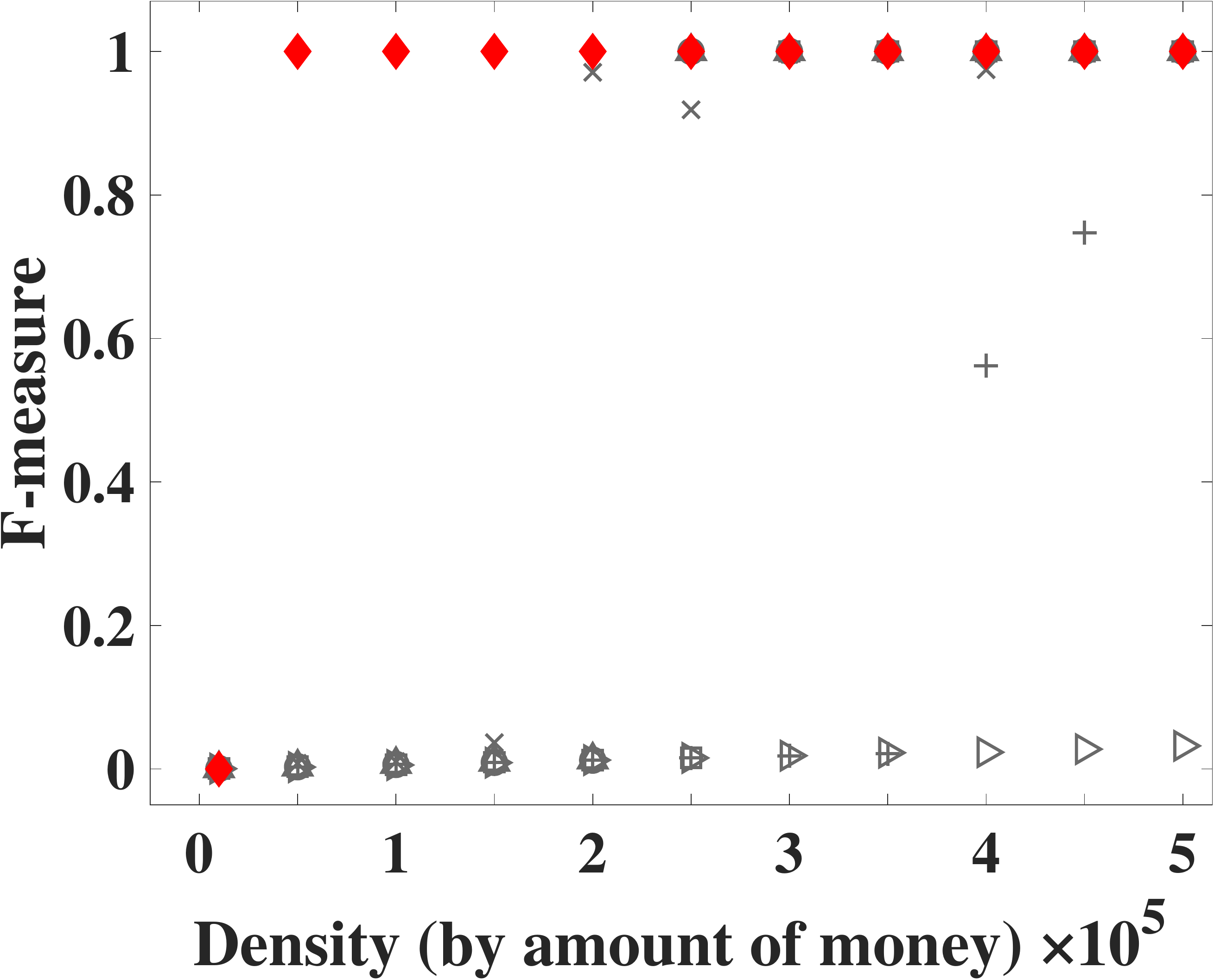}
        \end{minipage}
        \label{fig:cfd_dim3_inject1}
    }
    ~~~~
    \subfigure[ascending \# of accounts]{
        \begin{minipage}{0.36\linewidth}
            \includegraphics[width=\textwidth]{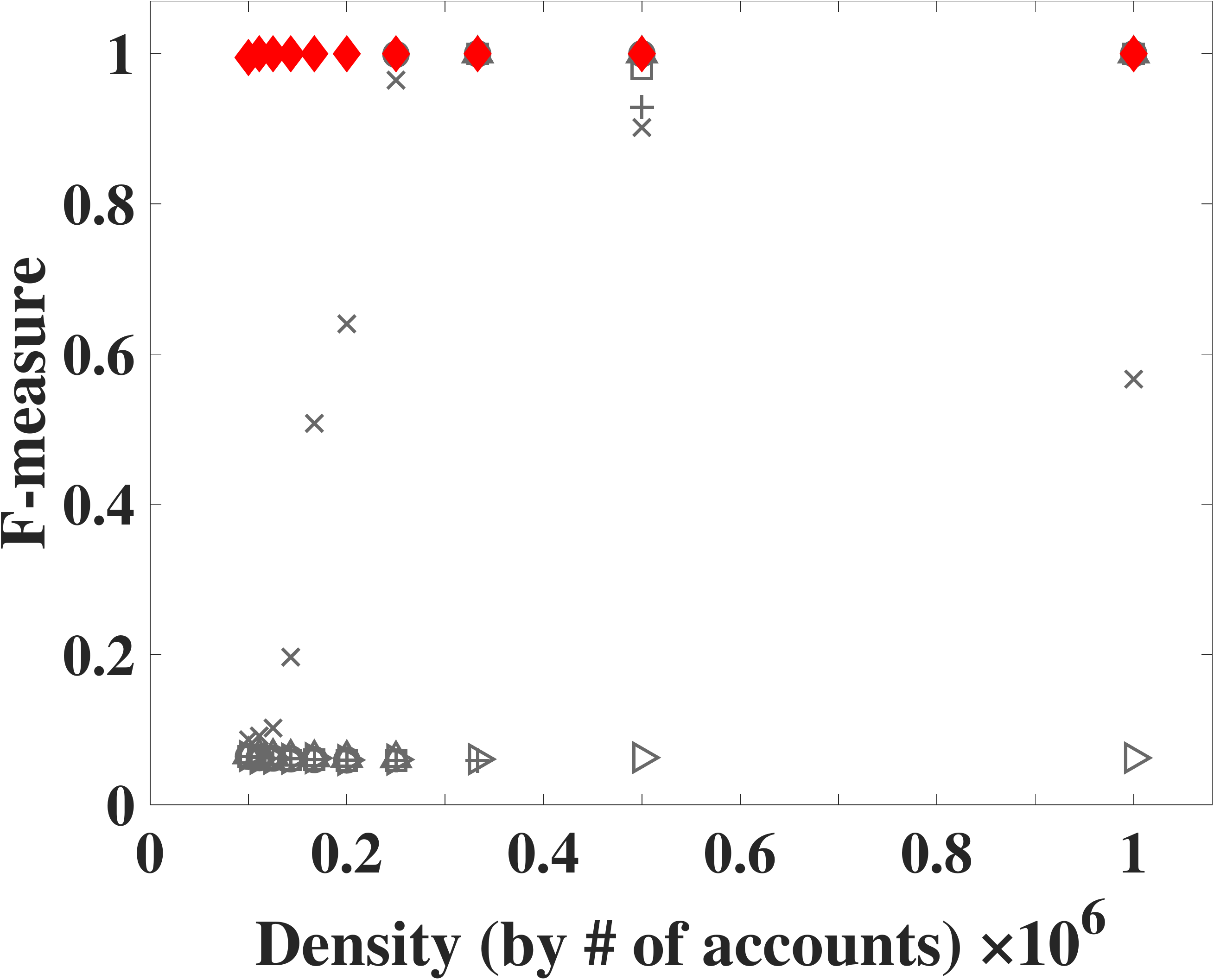}
        \end{minipage}
        \label{fig:cfd_dim3_inject2}
    }
    \caption{\method outperforms baselines under different injected densities by descending amount of money (a) or ascending \# of accounts (b) on CFD-$3$ data. }
    \label{fig:cfd_dim3_inject}
\end{figure}

\textbf{The influence of the amount of money:} 
In this experiment, the number of $B_x, B_y, B_z$ is $5$, $10$ and $5$ resp, and we increase the amount of injected money laundered step by step while fixing other conditions. As shown in Figure.~\ref{fig:cfd_dim3_inject1}, \method detects the ML behavior earliest and accurately, and the methods based on bipartite graph are unable to catch suspicious tripartite dense flows in the tensor.

\textbf{The influence of the number of fraudulent accounts:}
Another possible case is that fraudsters may employ as many as people to launder money, making the ML behavior much harder to detect. In this experiment, we increase the number of fraudsters step by step at a fixed ratio $(5:10:5)$ while keeping the amount of laundering money and other conditions unchanged. As Figure.~\ref{fig:cfd_dim3_inject2} shown, our method achieves the best results.

\textbf{Robustness with different injection ratios of accounts:}
To verify the robustness of our method, we randomly pick $B_x$, $B_y$ and $B_z$ under three ratios as presented in Table~\ref{tab:inject-radio}. The metric for comparison is \textbf{FAUC}: the areas under curve of F-measure as in Figure.~\ref{fig:cfd_dim3_inject}. We normalize the density in horizontal axis to scale FAUC between 0 and 1, and higher FAUC indicates better performance. And we can see from Table~\ref{tab:inject-radio}, \method achieves far better performance than other baselines under all settings, indicating earlier and more accurate detection for more fraudulent accounts.

\begin{table}[t]
    \centering
    \caption{Experimental results on CFD-$3$ data with different injection ratios of accounts}
  \scriptsize
    \begin{tabular}{c|c|c|c|c|c|c|c}
    \toprule
    \textbf{X:Y:Z} & \textbf{\method} & $\textbf{D-Cube}_{geo}$ & $\textbf{D-Cube}_{ari}$ & \textbf{CPD} & \textbf{CrossSpot} & $\textbf{M-Zoom}_{geo}$ & $\textbf{M-Zoom}_{ari}$ \\
    \midrule
    5:10:5 & \textbf{0.940} & 0.189 & 0.553 & 0.641 & 0.015 & 0.455 & 0.553\\
    \midrule
    10:10:10 & \textbf{0.940} & 0.427 & 0.653 & 0.647 & 0.024 & 0.555 & 0.555 \\
    \midrule
    10:5:10 & \textbf{0.970} & 0.652 & 0.652 & 0.725 & 0.020 & 0.652 & 0.652\\
    \midrule
    \end{tabular}
    \label{tab:inject-radio}
\end{table}

\begin{figure}[t]
	\centering
	\subfigure[CBank data with ground-truth]
	{\includegraphics[width=0.4\textwidth]{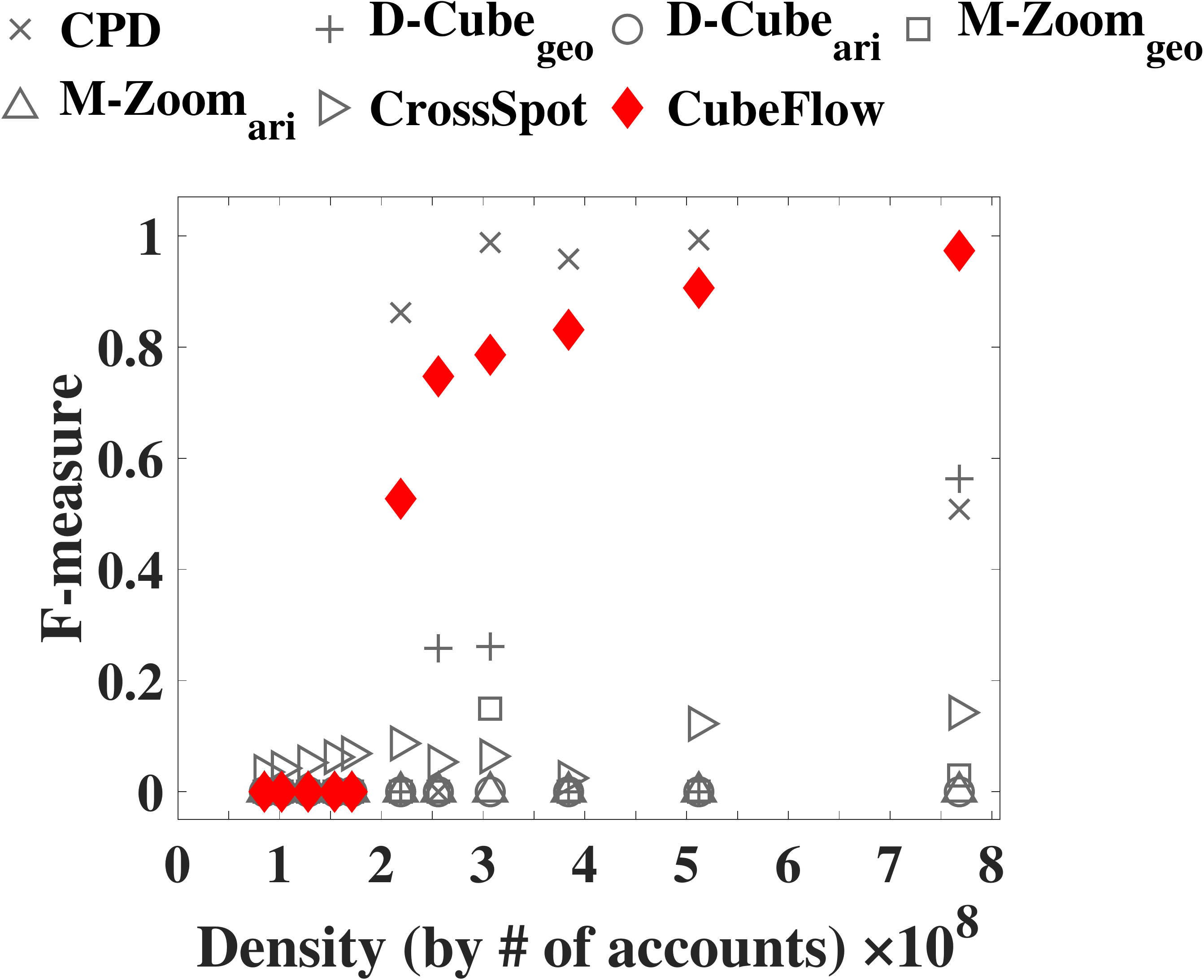}
	\label{fig:cbank_gt_density}
	}
	~~~~
	\subfigure[mass distribution of flows]
	{\includegraphics[width=0.35\textwidth]{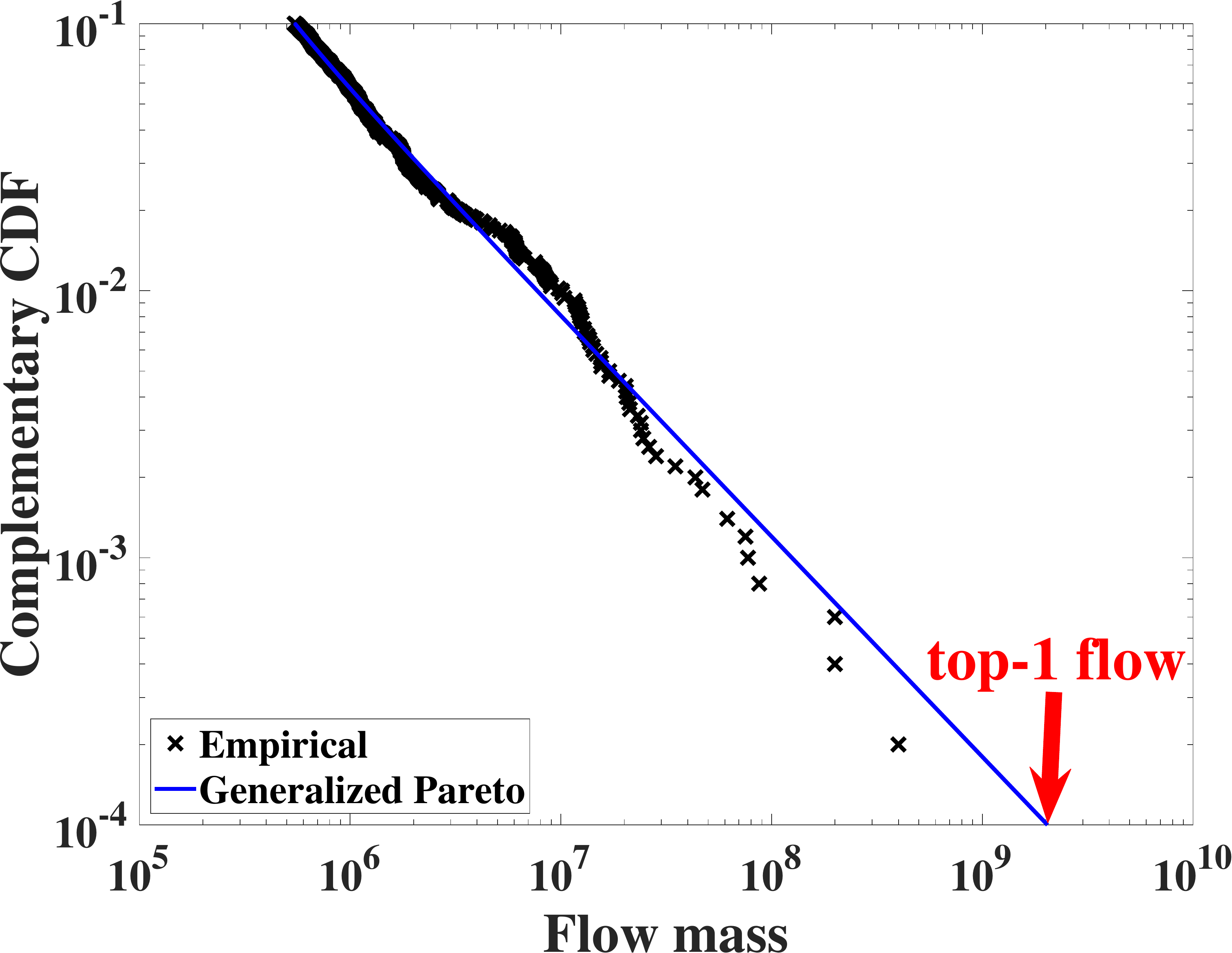}
	\label{fig:tail}}
	\caption{\method performs best in real CBank data. (a) \method detects earliest (less money being laundered) and accurately in ground-truth community. (b) The GP distribution closely fits mass distributions of real flows: Black crosses indicate the empirical mass distribution for flows with same size with top-$1$ flow detected by \method, in the form of its complementary CDF (i.e.~CCDF). }
\end{figure}

\subsection{Q2. Performance on real-world data}
CBank data contains labeled ML activity: based on  the  $X $ to $Y$ to  $Z$ schema, the number of each type of accounts is $4$, $12$, $2$.  
To test how accurately and early we can detect the fraudsters in CBank data, we first scale down the percentage of dirty money laundered from source accounts to destination accounts, then gradually increase the volume of money laundering linearly back to the actual value in the data.
Figure.~\ref{fig:cbank_gt_density} shows that \method can the catch the ML behaviors earliest, exhibiting our method's utility in detecting real-world ML activities. 
Note that although CPD works well at some densities, it fluctuate greatly, indicating that CPD is not very suitable for ML behavior detection.


\textbf{Flow Surprisingness estimation with extreme value theory:} 
 Inspired by \cite{hooi2020telltail}, we use Generalized Pareto (GP) Distribution, a commonly used probability distribution within extreme value theory, to estimate the extreme \textbf{tail} of a distribution without making strong assumptions about the distribution itself.
 GP distributions exhibit heavy-tailed decay (i.e.~power law tails), which can approximate the tails of almost any distribution, with error approaching zero~\cite{balkema1974residual}.
 
 Specifically, we estimate tail of GP distribution via sampling. Given a flow corresponding to two blocks, ${\mathbfcal{B}}_{\bf{P}}$, ${\mathbfcal{B}}_{\bf{Q}}$, with total mass $M({\mathbfcal{B}}_{\bf{P}})+M({\mathbfcal{B}}_{\bf{Q}})$, we sample $5000$ uniformly random flows from data with same size. For $\epsilon=0.1$, we fit a GP distribution using maximum likelihood to the largest $\epsilon N$ masses. The surprisingness of flow is the \textbf{CDF} of this GP distribution, evaluated at its mass. As shown in Figure.~\ref{fig:tail}, masses of sampled flows follow a GP distribution and tail measure score (i.e.~CDF) of top-$1$ flow detected by \method is very close to $1$ (pointed by red arrow), indicating that this activity is quite extreme(i.e.~rare) in CBank data.

\begin{figure}[t]
    \centering
    \begin{minipage}{0.8\linewidth}
        \includegraphics[width=\textwidth]{legend1row-crop}
    \end{minipage}
    \vfill 
    \subfigure[descending amount of money]{
        \begin{minipage}{0.36\linewidth}
            \includegraphics[width=\textwidth]{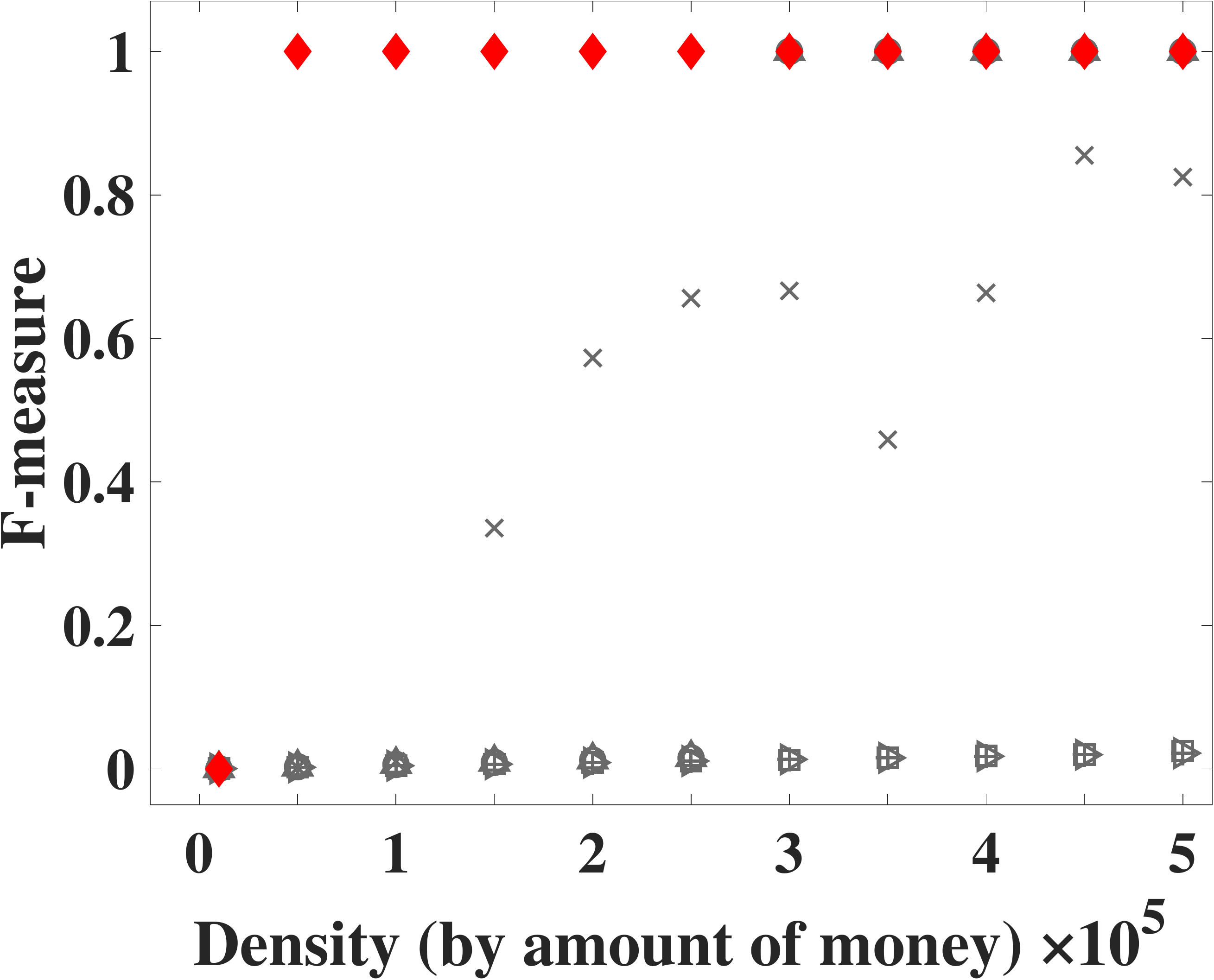}
        \end{minipage}
        \label{fig:cfd_dim4_inject1}
    }
    ~~~~
    \subfigure[ascending \# of accounts]{
        \begin{minipage}{0.36\linewidth}
            \includegraphics[width=\textwidth]{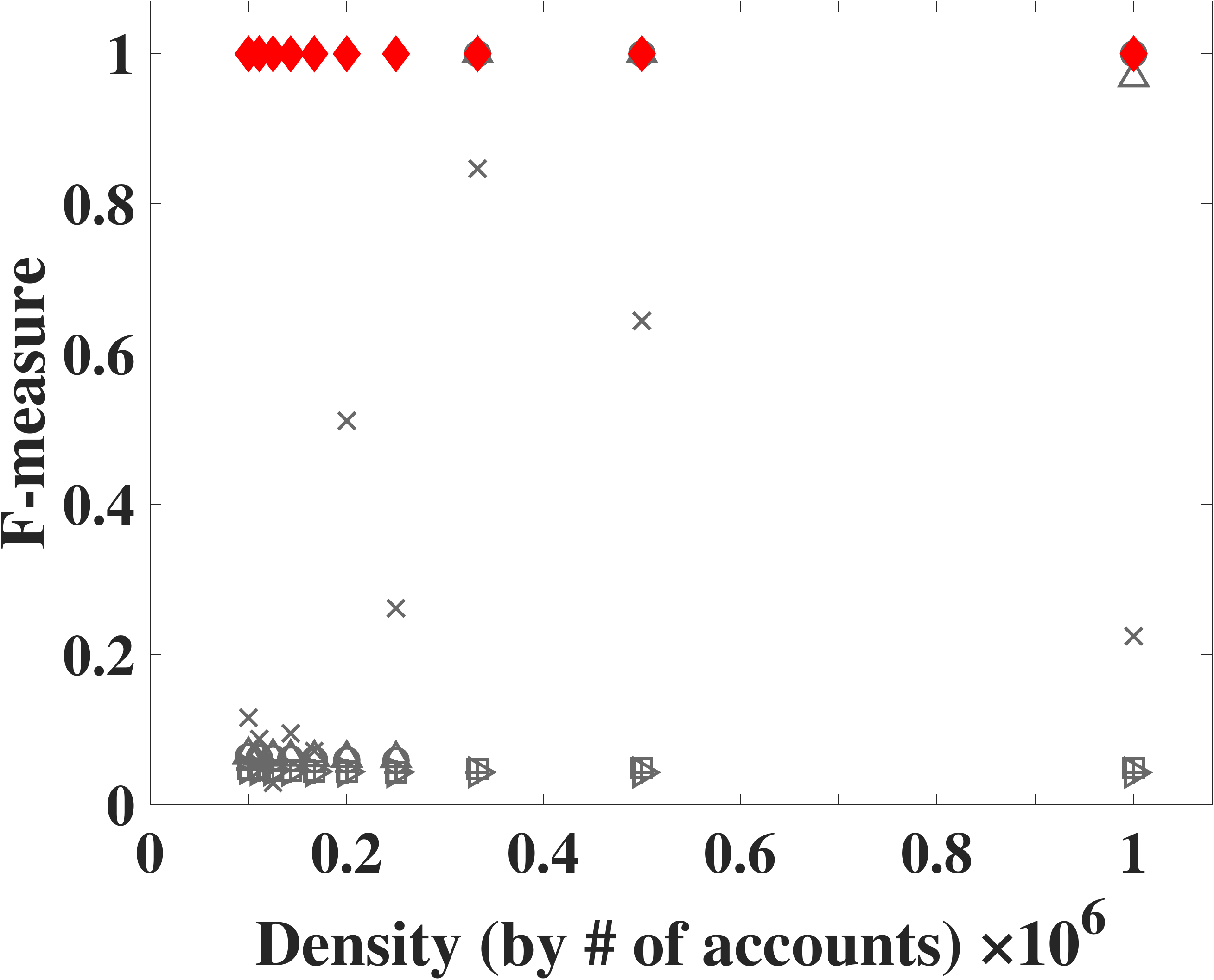}
        \end{minipage}
        \label{fig:cfd_dim4_inject2}
    }
    \caption{\method outperforms the baselines under different adversarial densities by descending the amount of money (a) or ascending \# of accounts (b) on CFD-$4$ data.}
    \label{fig:cfd_dim4_inject}
\end{figure}

\begin{figure}[t]
\centering
\includegraphics[width=0.36\textwidth]{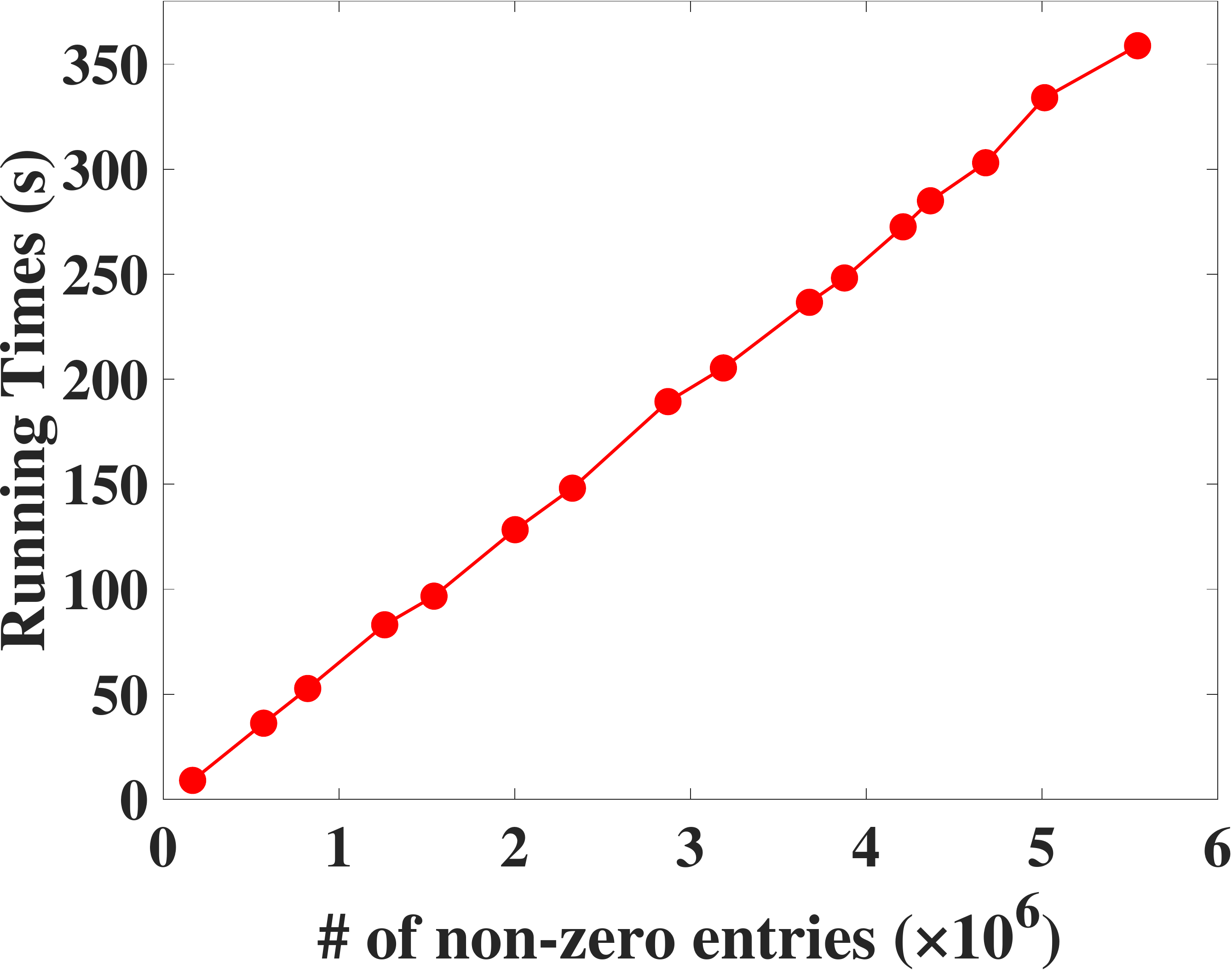}
\caption{\method scales near linearly}
\label{fig:gt_cost_time}
\end{figure}

\subsection{Q3. Performance on $4$-mode tensor}
To evaluate the performance of \method dealing with multi-mode data, we conducted similar experiments on CFD-$4$ data. Figure.~\ref{fig:cfd_dim4_inject1}-\ref{fig:cfd_dim4_inject2} show that our method takes significant advantages over the baselines as our method achieves excellent performance far earlier.

\subsection{Q4. Scalability}
%

\textbf{Scalability:} We demonstrate the linearly scalability with of \method by measuring how rapidly its update time 
increases as a tensor grows. As Figure.~{\ref{fig:gt_cost_time}} shown, \method scales linearly with the size of non-zero entries.

\section{Conclusion}
\label{sec:cons}
In this paper, we propose a money laundering detection method, \method, which is a scalable, flow-based approach to spot the fraud in big attributed transaction tensors. We model the problem with two coupled tensors and propose a novel multi-attribute metric which can utilize different characteristics of money-laundering flow. Experiments based on different data have demonstrated the effectiveness and robustness of \method's utility as it outperforms state-of-the-art baselines.
The source code is opened for reproducibility.


\subsubsection{Acknowledgements.}  
This paper is partially supported by the National Science
Foundation of China under Grant No.91746301, 61772498, U1911401, 61872206, 61802370. This paper is also supported by the
Strategic Priority Research Program of the Chinese Academy
of Sciences, Grant No. XDA19020400 and 2020 Tencent
Wechat Rhino-Bird Focused Research Program.


\end{document}